\documentclass[aps,pre,twocolumn,groupedaddress,showpacs,
amsmath,amssymb,amsfonts]{revtex4}

\usepackage{graphicx}
\usepackage{bm}
\usepackage{bbm}

\begin{document}

\title{Spectral correlations of individual quantum graphs}

\author{Sven Gnutzmann} \email[]{gnutz@physik.fu-berlin.de}
\affiliation{Institut f\"ur Theoretische Physik, Freie Universit\"at
  Berlin, Arnimallee 14, 14195 Berlin, Germany\\
  Department of Physics of Complex Systems, The Weizmann Institute of
  Science, Rehovot 76100, Israel
}

\author{Alexander Altland} \email[]{alexal@thp.uni-koeln.de}
\affiliation{Institut f\"ur Theoretische Physik, Universit\"at zu
  K\"oln, Z\"ulpicher Str.\ 77, 50937 K\"oln}

\date{\today}

\begin{abstract}
  We investigate the spectral properties of chaotic quantum graphs. We
  demonstrate that the `energy'--average over the spectrum of
  individual graphs can be traded for the functional average over a
  supersymmetric non--linear $\sigma$--model action. This proves that
  spectral correlations of individual quantum graphs behave according
  to the predictions of Wigner--Dyson random matrix theory. We explore
  the stability of the universal random matrix behavior with regard to
  perturbations, and discuss the crossover between different types of
  symmetries.
\end{abstract}

 \maketitle

\section{Introduction}

The spectral fluctuations of individual complex (chaotic) quantum
systems are universal and can be described in terms of Wigner--Dyson
random matrix theory\cite{Wigner,Dyson}. (See also
\cite{Mehta,Guhr,Haake,Stockmann,Efetov} and references therein).  For
classically chaotic systems this empirical statement was promoted to a
conjecture by Bohigas, Giannoni and Schmit~\cite{BGS}. (See also
\cite{Casati,Berry_anp}.) While, however, there is enormous
experimental and numerical evidence in support of this
conjecture~\cite{exceptions} the physical basis of universality is not
yet fully understood theoretically.

To date, the most advanced approach in developing correspondences
between spectral statistic and non--linear dynamics is semiclassical
analysis. Beginning with Berry's seminal work~\cite{Berry_diagonal} it
became understood that information on spectral correlations is stored
in action correlations of classical periodic orbits (see
also~\cite{action_correlations}.) Going beyond the `diagonal'
approximation~\cite{Berry_diagonal} wherein only identical (and
mutually time--reversed) orbits contributing to the Gutzwiller double
sum~\cite{Gutzwiller} are taken into account, a hierarchy of ever more
complex expansions in orbit pairs has been
constructed~\cite{SR,tau_2,tau_infinity}.  In this way, it was shown
that to all orders in an expansion in the ratio $\tau\equiv t/t_H$ the
short time ($t<t_H$) behavior of the spectral form factor $K(\tau)$ of
uniformly hyperbolic quantum systems coincides with the universal predictions of
random matrix theory (RMT). (Here, $t_H=\frac{2\pi\hbar}{\Delta E}$ denotes the Heisenberg
time and $\Delta E$ is the mean level spacing.) However, in view of
the fact that at $\tau=1$, the function $K(\tau)$ contains an
essential singularity, it is presently not clear how to extend this
expansion to times larger than the Heisenberg time.

Some time ago, a field theoretical approach to quantum chaos ---
dubbed the ballistic $\sigma$--model --- has been
introduced \cite{AAA} as an alternative to semiclassical expansions. The most
promising aspect of this development is that in field theory the full
information on universal RMT correlations is obtained in a very simple
manner, viz. by integration over globally uniform `mean field'
configurations; universality of chaotic quantum systems is proven,
once it has been shown that at sufficiently low energies (long times)
fluctuations become negligible and the field theory indeed reduces to
its mean field sector. Unfortunately, however, it has so far not been
possible to demonstrate this reduction in a truly convincing
manner. 
(The situation is much better in the field of disordered
chaotic systems: It has been known for some time that at low energies disordered
systems exhibit RMT spectral correlations upon configurational
averaging. This type of universality has been proven~\cite{Efetov} by
field theoretical methods similar to those mentioned above.)

Motivated by the lack of universality proofs for generic quantum
systems with underlying Hamiltonian chaos, we have recently considered
the spectral properties of quantum graphs~\cite{GA}. 
(For the general theory of
quantum graphs, see \cite{Kuchment} and references therein.) Quantum
graphs differ from generic Hamiltonian systems in two crucial aspects:
First, the classical dynamics on the graph is not deterministic.  It
is rather described by a Markov process.  Second, quantum graphs are
`semiclassically exact' in that their spectrum can be exactly described
in terms of trace formulae. In spite of these differences, quantum
graphs display much of the behavior of generic hyperbolic quantum
systems~\cite{Kottos} (while being not quite as defiant to analytical
treatment than these.)

Earlier work on universal spectral statistics in quantum graphs was
based on periodic orbit summation schemes similar in spirit to the
semiclassical approach to Hamiltonian systems. Specifically,
Berkolaiko {\it et al.}~\cite{greg} developed a perturbative
diagrammatic language to analyze the periodic--orbit expansions of
spectral correlation functions beyond the diagonal
approximation. Tanner~\cite{Tanner} analyzed the structure of the
semiclassical expansion to conjecture criteria for the presence of
universal correlations on graphs. He also established connections
between universality and the decay rates of classical Markovian
dynamics of the system (for details see appendix~\ref{sec:tanner}).

While all building blocks of semiclassical analysis on graphs are
known~\cite{greg,GS}, and a complete summation over all orbit pairs
may be in reach, semiclassics on graphs is subject to the same
limitations as in Hamiltonian systems. In particular, it is not clear
how to extend its domain of applicability to times beyond the
Heisenberg time. In view of these difficulties, we have developed an
alternative approach~\cite{GA} which is based on field theoretical
methods and avoids diagrammatic resummations
altogether. Rather, it is based on two alternative pieces of input,
both of which have been discussed separately before:
\begin{itemize}
\item[\textit{i.}]  The exact equivalence of a spectral average for a
  quantum graph with incommensurate bond lengths to an average over a
  certain ensemble of unitary matrices \cite{Kottos,Barra,Tanner}.
\item[\textit{ii.}] An exact mapping of the phase--averaged spectral
  correlation functions onto a variant of the supersymmetric
  $\sigma$--model by an integral transform known as the color--flavor
  transformation~\cite{Zirnbauer}.
\end{itemize}
The synthesis of \textit{i.} and \textit{ii.}~\cite{GA} leads to a
formulation similar in spirit to the `ballistic $\sigma$--model' yet
not burdened by the technical problems of that approach.  It is  the
purpose of this paper to give a detailed account of this theory, and
to discuss a number of generalizations. Specifically, we will discuss
the crossover between systems of conserved (orthogonal symmetry) and
broken (unitary symmetry) time--reversal invariance, and we will
consider the case of broken spin rotation invariance (symplectic
symmetry.)

The paper is organized as follows: In Section \ref{sec:graphs} we give
a short introduction to quantum graphs. We discuss the relevant
quantization conditions, spectral correlators, and the
meaning of incommensurate bond lengths. The supersymmetry approach to
the spectral two--point correlation function is discussed in
Section~\ref{sec:susyrepresentation}.  In Section
\ref{sec:saddlepoint} we subject the supersymmetric generating
functional to a stationary phase analysis. We show under which
conditions the field theory can be reduced to a `mean field' theory
of RMT--type correlations. We also discuss the crossover between
graphs of orthogonal and unitary symmetry. Quantum graphs belonging to
the symplectic symmetry class are discussed in appendix \ref{sec:GSE},
and an outline of the proof of the color--flavor transformation is
given in appendix~\ref{sec:color-flav-transf}.

\section{Quantum graphs \label{sec:graphs}}

\subsection{Generalities}
\label{sec:generalities}

\begin{figure}[hbt]
  \includegraphics[width=9cm]{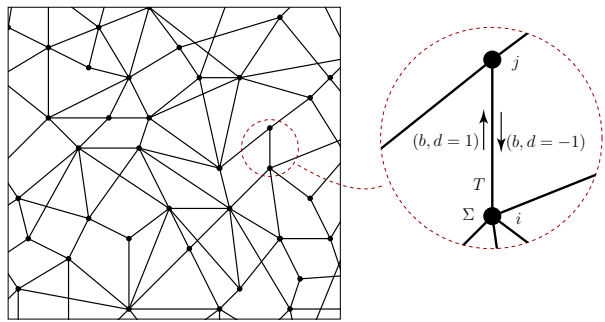}
  \caption{Schematic of a fraction of a (planar) simple quantum graph
    and of the notation used in the text.}
  \label{fig:quantum_graph}
\end{figure}

A finite graph ${\mathcal{G}}$ consists of $V$ vertices which are
connected by $B$ bonds.  The $V\times V$ connectivity matrix is
defined by
\begin{equation}
  C_{i,j}= \# \left\{\textrm{bonds
    connecting the vertices $i$ and $j$}\right\}  .
\end{equation}
A graph is simple if for all $i,\ j$, $C_{i,j}\in [0,1]$ (no parallel
connections) and $C_{i,i}=0$ (no loops).  The number of bonds is
$B=\frac{1}{2} \sum_{i,j=1} ^V C_{i,j} $.  The valency of a vertex $i$
is the number of bonds connected to it $v_i = \sum_{j=1} ^V C_{i,j}$.
A graph is called `connected' if it cannot be split into disjoint
sub--graphs. With only slight loss of generality~\cite{fn:simple}, we
will focus on the case of simple connected graphs throughout (cf.
Fig.~\ref{fig:quantum_graph} for a schematic.)

We denote a bond connecting the vertices $i$ and $j$ by $b=(ij)$.
The notation $(ij)$ and the letter $b$ will be used whenever we refer
to bonds without specifying a direction: $b\in (ij)=(ji)$. A directed
bond $\beta=(b,d)$ consists of a bond $b$ and a direction on the bond
which will be denoted by a direction index $d=\pm 1$.  For $b= (ij)$
and $i<j$ we set $d=+1$ for the direction $i \rightarrow j$ and $d=-1$
on the opposite direction.

The position $x$ of a point on the graph is determined by specifying
its bond $b$, and its distance $x_b\in [0,L_b]$ from the adjacent
vertex with the smaller index. The length of a bond is denoted by
$L_b$.  Throughout, we will assume the bond lengths to be
\emph{incommensurable} (or rationally independent) in the sense that
there is no non--vanishing set of integers $m_b \in \mathbbm{Z}$ such
that $\sum_b m_b L_b=0$.

The Schr\"odinger operator on
${\mathcal{G}}$ is defined by one--dimensional Laplacians on the
bonds, and a set of vertex boundary conditions establishing
self--adjointness.  Its wave functions $\Psi(x)$ are complex valued,
piecewise continuous and bounded functions.  Writing $\Psi(x)\equiv
\psi_b(x_b)$ for $x=x_b\in(0,L_b)$, the solutions of the stationary
Schr\"odinger equation at a given wave number $k>0$ have the form
\begin{equation}
  \begin{split}
    &\psi_b(x_b;k)=e^{-i A_b(x_b-\frac{L_b}{2})}\times\\
    &\qquad \left( \alpha_{b,+1} e^{ik(x_b-\frac{L_b}{2})}+
      \alpha_{b,-1} e^{-ik(x_b-\frac{L_b}{2})} \right),
  \end{split}
  \label{eq:bond-wf}
\end{equation}  
where $\alpha_{b,d=\pm 1}$ are the complex amplitudes of `right'
($d=+1$) and `left' ($d=-1$) propagating waves on the bond, and $A_b$
are constant phases generated by optional magnetic fluxes threading
the plaquettes of the graph.  To characterize the vertex boundary
conditions, we introduce ($k$--independent) $v_i \times v_i$ vertex
scattering matrices $\Sigma^{(i)}_{b,b'}$ connecting incoming waves to
outgoing waves at $i$. ($b$ and $b'$ run over bonds connected to $i$).
Denoting the outgoing/incoming direction on bond $b$ by
$d_{\mathrm{out}}$/$d_{\mathrm{in}}$, these matrices are defined by
the equation
\begin{equation}
  \begin{split}
    &\alpha_{bd_{\mathrm{out}}}
    e^{-i(k-d_\mathrm{out}A_b)\frac{L_b}{2}}=\\
    &\qquad \qquad \sum_{b'} \Sigma^{(i)}_{b,b'} e^{i (k-d_\mathrm{in}
      A_{b'}) \frac{L_{b'}}{2}} \alpha_{b'd_\mathrm{in}},
  \end{split}
  \label{eq:bc1}
\end{equation}
To represent this equation in a more concise form, we combine all
amplitudes $\alpha_{b d}$ into a $2B$-dimensional vector
$\vec{\alpha}$. In this notation,
\begin{equation}
  \vec{\alpha}= \mathcal{U}_B(k) \vec{\alpha}
  \label{eq:bc2}
\end{equation}
where the $2B\times 2B$ quantum map~\cite{bondscattering} is given by
\begin{equation}
  \mathcal{U}_B(k)= T(k) S T(k),
\end{equation}
the diagonal matrix $T(k)_{b\,d, b'\,d'}=\delta_{b\,d, b'\,d'} e^{i
  (k+d A_b) L_b/2}$ describes the propagation along half a bond, and
\begin{equation}
  S_{bd, b'd'}=
  \begin{cases}
    \Sigma^{(i)}_{b, b'} &
    \text{$(b',d') \rightarrow i \rightarrow (b,d)$} \\
    0 & \text{else},
  \end{cases}
\end{equation}
combines all boundary conditions at the vertices into a single
scattering matrix. The equivalent of the quantum map in a Hamiltonian
system is a quantized Poincar\'e map.

The boundary condition \eqref{eq:bc2} can be fulfilled only for
discrete set of wave numbers $k_n$. These numbers define the spectrum
of the quantum graph. For $k>0$ the quantization condition
\eqref{eq:bc2} is equivalent to the vanishing of the spectral
determinant
\begin{equation}
  \label{xi_det}
  \xi(k)\equiv\mathrm{det}\left(\mathbbm{1} -
    \mathcal{U}_B(k)\right).
\end{equation}
Thus $\xi(k)=0$ if and only if $k=k_n>0$ is in the spectrum.

By way of
example, we mention two frequently employed families of boundary
conditions: so--called \emph{Neumann boundary
  conditions}~\cite{neumann} correspond to
\begin{equation}
  \Sigma^{(i),\mathrm{N}}_{b,b'}=
  \frac{2-\delta_{b,b'} v_i}{v_i}.
  \label{eq:neumann_bc}
\end{equation}
For large valencies $v_i$ the non-diagonal terms are much smaller than
the diagonal, and the back scattering term dominates. While on general
graphs wave functions need not be continuous across the vertices, they
are so on Neumann graphs~\cite{Kottos}.  Another interesting set of
boundary conditions is implemented through {\it discrete Fourier
  transform} (DFT) matrices
\begin{equation}
  \Sigma^{(i),\mathrm{DFT}}_{b,b'}=\frac{1}{\sqrt{v_i}}
  \exp\left(2\pi i \frac{\hat\pi(b) \hat\pi (b')}{v_i}\right)
\end{equation}
where $b \mapsto \hat\pi (b)$ maps the $v_i$ bonds connected to vertex
$i$ one-to-one onto the numbers $1,2\dots,v_i$. These boundary
conditions do not imply continuity at the vertices; incoming wave
packets are scattered into the outgoing bonds with equal probability.

\subsection{Time--reversal invariance \label{sec:timerev}}

As with Hamiltonian chaotic symmetries, quantum graphs of different
symmetries may be identified. Specifically, quantum graphs carrying
spin degrees of freedom (and spin--rotation invariance breaking vertex
scattering matrices) fall into the symplectic or unitary symmetry
class depending on whether time reversal invariance is broken or not.
These cases will be discussed in appendix~\ref{sec:GSE}. In the
absence of spin, we need to distinguish between graphs with broken
(unitary symmetry or symmetry class $A$ in the notation of~\cite{AZ})
or conserved (orthogonal symmetry or symmetry class $A$I) time
reversal invariance.

A quantum system is time--reversal invariant if its Hamiltonian $H$
commutes with an anti--unitary time--reversal operator $\mathcal{T}$,
$\left[ H,\mathcal{T}\right]=0$. For spinless systems, $\mathcal{T}$
is an involutory operator, $\mathcal{T}^2= \mathbbm{1}$\cite{Haake}.
The condition $\left[H,\mathcal{T}\right]=0$ restricts the form of
both the bond propagation matrix $T(k)$ and the vertex scattering
matrices $\Sigma(k)$. In non--time reversal invariant systems, these
matrices obey no conditions other than unitarity. However, for
conserved time reversal invariance, and with a definitive choice of
the time--reversal operator $\mathcal{T}$, all vertex scattering
matrices have to be symmetric $\Sigma^{(i)}=\Sigma^{(i)\; T}$, i.e.
\begin{equation}
  S=S^\mathcal{T}\equiv 
  \sigma^\mathbf{dir}_\mathrm{1}
  S^T \sigma^\mathbf{dir}_\mathrm{1}
  \label{eq:timerev1}
\end{equation}
where $\sigma^\mathbf{dir}_\mathrm{1}= \left(
  \begin{smallmatrix}
    0 & \mathbbm{1}\\
    \mathbbm{1} & 0
  \end{smallmatrix}
\right)$ is the Pauli matrix in direction indices $d$. Additionally,
all magnetic phases $A_b$ must vanish. The crossover between
orthogonal and unitary symmetry will be discussed
in~\ref{sec:crossover} where we explore the consequences of a gradual
switching on of magnetic phase factors.

\subsection{The density of states and 
  spectral correlation functions of quantum graphs \label{sec:dos}}

The density of states (DoS) of a quantum graph is defined as
\begin{equation}
  d(k)= \sum_n \delta(k-k_n)=\frac{1}{\Delta}
  +\delta d(k)
\end{equation}
where the sum runs over the spectrum $k_n$.  We have written the DoS
as a sum over a smooth part $1/\Delta$ where $\Delta$ is the mean
level spacing and fluctuations $\delta d(k)$. Both parts allow for an
explicit representation in terms of the quantum evolution map.  For
the mean (or Weyl) part one obtains
\begin{equation}
  \frac{1}{\Delta}=\frac{1}{2 \pi i}
  \frac{d}{dk} \mathrm{ln}\,
  \mathrm{det}\left(-\mathcal{U}_B(k)\right)=
  \frac{B \overline{L}}{\pi},
\end{equation}
where $\overline{L}=\frac{1}{B}\sum_b L_b$ is the mean bond length.
Note that the mean level spacing is constant.  The fluctuations can be
expressed through the spectral determinant\cite{Kottos}
\begin{equation}
  \delta d(k) 
  = -
  \frac{1}{\pi} 
  \mathrm{Im}\, \frac{d}{dk} \mathrm{ln}\,
  \xi(k^+),
\end{equation}
where $k^+\equiv k+i\epsilon$ and the limit $\epsilon \rightarrow 0$
is implied.  Using that $\mathrm{ln}\, \mathrm{det}\, A =
\mathrm{tr}\, \mathrm{ln}\, A$ and expanding the logarithm one obtains
an \emph{exact} Gutzwiller type trace formula
\begin{equation}
  \delta d(k)=
  \frac{1}{\pi} 
  \mathrm{Im} \frac{d}{dk}
  \sum_{n=1}^\infty \frac{1}{n} \mathrm{tr\,} 
  \mathcal{U}_B^n(k^+)
\end{equation}
expressing the DoS in terms of a sum over periodic orbits (periodic
sequences of directed bonds.)

We aim to explore the statistical properties of the fluctuating part
of the DoS.  The $N$-point DoS correlation function is defined by an
average over the complete spectrum
\begin{equation}
  \begin{split}
    &R_N(s_1,\dots,s_{N-1})\equiv\\
    &\Delta^N \left\langle \delta d(k+s_{N-1}\Delta)\dots
      d(k+s_1\Delta)\delta d(k)\right\rangle_k,
  \end{split}
\end{equation} 
where
\begin{equation}
  \langle f(k) \rangle_k \equiv \lim_{K\rightarrow \infty}
  \frac{1}{K} \int_0^K dk f(k).
\end{equation}
Throughout, we will focus attention on the two--point correlation
function $R_2(s)$. The two--point function can be conveniently
expressed as a derivative of quotients of spectral determinants:
\begin{equation}
\label{R2_specdet}
  R_2(s)=\frac{1}{8 g^2 \pi^2}\frac{d^2}{dj_+ dj_-}
  \big|_{j=0}\mathrm{Re}\, 
  \left\langle \zeta(j_+,j_-;s)\right\rangle_k,
\end{equation}
where
\begin{equation}
  \zeta(j_+,j_-;s) \equiv 
  \frac{\xi(k^+ +p_{+\mathrm{\bf f}})}{
    \xi(k^+ +p_{+\mathrm{\bf b}})}
  \left(
    \frac{\xi(k^+ +p_{-\mathrm{\bf f}})}{ 
      \xi(k^+ +p_{-\mathrm{\bf b}})}
  \right)^*
\end{equation}
and $p_{\pm\mathrm{\bf b}}=(\pm s/2 - j_\pm)\Delta$,
$p_{\pm\mathrm{\bf f}}=(\pm s/2 + j_\pm)\Delta$. (Higher order
correlation functions may be obtained in a similar manner from
generating functions involving additional quotients of spectral
determinants.)

For later reference, we recall that the RMT two--point correlation
functions are given by
\begin{equation}
  \begin{split}
    R_2^\mathrm{GUE}(s)=& \delta(s)- \frac{\sin^2 \pi s }{\pi^2 s^2}
    \\
    R_2^\mathrm{GOE}(s) =&
    R_2^\mathrm{GUE}(s)+\\
    & \frac{\left(\pi |s| \cos \pi s - \sin \pi |s|\right) \left(2\,
        \mathrm{Si}(\pi |s|)-\pi\right) }{2 \pi^2 s^2},
  \end{split}
\end{equation}
where $\mathrm{Si}(x)=\int_0^x dx' \, \frac{\sin x'}{x'}$ is the
sine integral.

We also notice that the statistical properties of the graph may be
characterized by correlation functions different yet closely allied to
the correlation functions introduced above: for any value of $k$ the
quantum map $\mathcal{U}_B$ possess a set of $2B$ `eigenphases' $e^{-i
  \theta_l(k)}$ ($l=1,\dots,2B$) on the unit-circle. At fixed $k$ the
density of phases is given by
$$
\rho_k(\theta) = \sum_{l=1}^{2B} \delta_{2\pi}(\theta-\theta_l(k))
$$
where $\delta_{2\pi}(\theta)$ is the $2\pi$-periodic
delta--function.  The statistical properties of this quantity are
defined by averaging over both $\theta$ and $k$.  Occasionally ---
e.g. within the context of the periodic orbit approach to graphs ---
it is sometimes advantageous to consider the correlation functions of
the eigenphases instead of the spectral correlators introduced above.
Under mild conditions (weak fluctuations in the bond lengths) both
types of correlators are equivalent in the limit $B\rightarrow \infty$
of large graphs. In the following we will keep our discussion focused
on the spectral correlators. With small and straight forward changes
our theory can be applied to the eigenphase-correlations as well.

\subsection{Consequences of incommensurability
  \label{sec:phaseaverage}}

The quantum map $\mathcal{U}_B(k)=T(k)S T(k)$ depends on the wave
number $k$ via the $B$ diagonal elements $e^{i k L_b/2}$.  Defining
$\phi_b(k)\equiv k L_b/2$ we have a map $k\mapsto (e^{i
  \phi_1(k)},\dots,e^{i \phi_B(k)})$ of the wavenumber into a
$B$-torus $T^B \subset \mathbbm{C}^B$. This map may be interpreted as
a `Hamiltonian flow' where $k$ plays the role of `time'.  As we assume
incommensurable bond lengths $L_b$, the image of the phase map covers
the torus densely, i.e. the Hamiltonian flow is `ergodic'. This in
turn means that long time averages ($k$--averages) may be traded for
phase space averages (averages over the torus or, equivalently,
independent averages over the $B$ phases $\phi_b$~\cite{Barra}):
\begin{equation}
\label{ave_equi}
  \begin{split}
    \Big\langle f \big( \{ e^{i \phi_b(k)} \} \big) \Big\rangle_k =&
    \Big\langle f \big( \{ e^{i \phi_b} \} \big)
    \Big\rangle_\phi\\
    \equiv& \frac{1}{(2\pi)^B} \int_{T^B} d^B \phi\, f \big( \{ e^{i
      \phi_b} \} \big).
  \end{split}
\end{equation}
It is this equivalence which makes the analytical calculation of
spectral correlation functions a feasible task. Upon replacing
$\langle\;\rangle_k \to \langle\;\rangle_\phi$, the one--parameter
family of matrices $T(k)\to T(\{\phi_b\})$ becomes an 'ensemble' of
random matrices. There is a well developed analytical machinery
designed to perform random phase averages of this kind. Below, we will
apply the formalism of supersymmetry to compute the random phase
averaged spectral correlation functions of the graph which, by virtue
of the equivalence above, are strictly equivalent to the wave number
averaged correlation functions.
 
\section{Nonlinear $\sigma$ model  
  for quantum graphs 
  \label{sec:susyrepresentation}} 

Consider the representation \eqref{R2_specdet} of the two--point
correlation function in terms of a double derivative of the quotient
$\zeta$ of spectral determinants. Replacing the $k$--average by a
random phase average, $\langle\zeta \rangle_k \to \langle \zeta
\rangle_\phi$, it is the purpose of the present section to derive a
$\sigma$--model representation of the two--point correlation function.

\subsection{The generating function as a Gaussian superintegral} 

Defining the supervectors
\begin{equation}
  \psi=
  \begin{pmatrix}
    s_1\\
    \dots\\
    s_N\\
    \chi_1\\
    \dots\\
    \chi_N
  \end{pmatrix}
  \;\;
  \text{and}
  \;\;
  \tilde\psi=
  \begin{pmatrix}
    s_1^* & \dots & s_N^* & \tilde\chi_1 & \dots \tilde\chi_N
  \end{pmatrix},
\end{equation}
where $s_i$ are complex commuting variables while $\chi_i$ and
$\tilde\chi_i$ are independent anti-commuting numbers, the quotient of
determinants of an $N \times N$ matrix $A_\mathbf{f}$ and a (positive)
$N \times N$ matrix $A_\mathbf{b}$ can be represented as a Gaussian
integral
\begin{equation}
  \frac{\mathrm{det} \, A_\mathbf{f}}{
    \mathrm{det}\, A_\mathbf{b}}\equiv \mathrm{sdet}\, A=
  \int d(\tilde\psi,\psi) e^{- \tilde\psi A \psi}.
\end{equation}
Here,
\begin{equation}
  A=
  \begin{pmatrix}
    A_\mathbf{b} & 0\\
    0 & A_\mathbf{f}
  \end{pmatrix}
\end{equation}
is a block--matrix in boson--fermion space (the two component space
introduced by the $s/\chi$ grading of $\psi$) and the measure is given
by
\begin{equation}
  d(\tilde\psi,\psi)= \pi^{-N}\prod_{i=1}^N 
  d\mathrm{Re}(s_i) d\mathrm{Im}(s_i) 
    d\tilde\chi_i d\chi_i
\end{equation}
where $\int d\chi_i \, \chi_i=1$ and $\int d\chi_i=0$.  We wish to
apply this relation to represent the spectral determinants
\eqref{xi_det} in terms of Gaussian integrals. In view of our
applications below, it will be convenient
to double the matrix dimensions~\cite{double_dim} using
\begin{equation}
  \xi(k+p)=\mathrm{det}\left(\mathcal{U}_B(p)\right)
  \mathrm{det}
  \begin{pmatrix}
    \mathbbm{1} & T(k)\\
    T(k) & \mathcal{U}_B(p)^\dagger
  \end{pmatrix}
\end{equation}
which leads to
\begin{equation}
  \zeta(j_+,j_-;s)=\int d(\tilde\psi,\psi)
  (\mathrm{sdet}(T_+ T_-^\dagger))^2
  e^{-{\bm{S}}[\tilde\psi,\psi]}
\end{equation}
where
\begin{equation}
  \begin{split}
    \bm{S}[\tilde\psi,\psi]=& \tilde\psi_+
    \begin{pmatrix}
      1 & T(k)\\
      T(k) &(T_+S T_+)^\dagger
    \end{pmatrix}
    \psi_+
    +\\
    & \tilde\psi_-
    \begin{pmatrix}
      1 & T(k)^\dagger\\
      T(k)^\dagger & T_-S T_-
    \end{pmatrix}\psi_-.
  \end{split}
  \label{eq:action1}
\end{equation}
Here, $\psi=\{\psi_{a,s,x,d,b}\}$ is a $16B$-dimensional supervector
where, $a=\pm$ distinguishes between the retarded and the advanced
sector of the theory (components coupling to $\xi$ or $\xi^*$,
respectively).  The index $s=\mathrm{\bf f},\mathrm{\bf b}$ refers to
complex commuting and anti--commuting components (determinants in the
denominator and numerator, respectively), and $x=1,2$ to the internal
structure of the matrix kernel appearing in \eqref{eq:action1}.  The
matrices
\begin{equation}
  T_\pm \equiv
  \begin{pmatrix}
    T(p_{\pm,\mathbf{b}}) & 0\\
    0 & T(p_{\pm,\mathbf{f}})
  \end{pmatrix}
\end{equation}
are diagonal matrices in superspace containing the appropriate bond
matrices $T$ in the boson--boson/fermion-fermion sector.

To account for the (optional) 
time-reversal invariance of the scattering matrix,
we introduce the doublets
\begin{equation}
  \begin{split}
    \Psi=& \frac{1}{\sqrt{2}}
    \begin{pmatrix}
      \psi\\
      \sigma_1^\mathbf{dir}\tilde\psi^T
    \end{pmatrix}
    \\
    \tilde\Psi=& \frac{1}{\sqrt{2}}
    \begin{pmatrix}
      \tilde\psi, & \psi^T \sigma_1^\mathbf{dir} \sigma_3^\mathbf{BF}
    \end{pmatrix},
  \end{split}
\end{equation}
where $\sigma_3^\mathbf{BF}\equiv \left(
\begin{smallmatrix}
  \mathbbm{1} & 0\\
  0 & -\mathbbm{1}
\end{smallmatrix}
\right)$ is the Pauli matrix in superspace. Notice that the lower
components of $\Psi$ emanate from the upper component by a time
reversal operations (transposition followed by inversion in
directional space.) For later reference, we note that new fields
depend on each other through the generalized transposition
\begin{equation}
  \Psi=\tau \tilde\Psi^T 
  \qquad 
  \tilde{\Psi}=\Psi^T \tau.
\end{equation}
The explicit definition of the matrix $\tau$ is given by
\begin{equation}
\label{tau_def}
  \begin{split}
    \tau=&\sigma_1^\mathbf{dir} \tau_0
    \\
    \tau_0=& E_\mathbf{b}\sigma_1^\mathbf{tr}
    -iE_\mathbf{f}\sigma_2^\mathbf{tr},
  \end{split}
\end{equation}
where $\sigma_i^\mathbf{tr}$ are Pauli matrices in the newly
introduced `time-reversal' space and $E_\mathbf{b/f}$ are the
projectors on the bosonic/fermionic sectors. However, all we will need
to know to proceed is that $\tau$ obeys the conditions
\begin{equation}
\label{tau_prop}
  \tau^T=\tau^{-1}
  \qquad  \text{and}
  \qquad 
  \tau^2= \sigma_3^\mathbf{BF}.
\end{equation}
The appearance of the matrix $\tau$ in conjunction with a
transposition operation suggests to introduce the generalized matrix
transposition
\begin{equation}
  A^\tau \equiv \tau A^T \tau^{-1}.
\end{equation}
Using Eq.~\eqref{tau_prop} and that~\cite{Efetov}
$(A^T)^T=\sigma_3^\mathbf{BF} A \sigma_3^\mathbf{BF}$, one finds that
the generalized transposition in involutory,
\begin{equation}
  (A^\tau)^\tau=A.
\end{equation}
For later reference we also note that
\begin{equation}
  \tilde{\Psi}_+ 
  A 
  \Psi_- 
  =
  \Psi_-^T 
  \sigma_3^\mathbf{BF} 
  A^T 
  \tilde{\Psi}^T_+
  =
  \tilde{\Psi}_- 
  A^\tau 
  \Psi_+.
\end{equation}
With all these definitions, the action \eqref{eq:action1} now takes
the form
\begin{equation} 
  \begin{split}
    \bm{S}[\tilde\Psi,\Psi]=& \tilde\Psi_+
    \begin{pmatrix}
      1 & T(k)\\
      T(k) &\mathcal{S}_+^\dagger
    \end{pmatrix}
    \Psi_+
    +\\
    & \tilde\Psi_-
    \begin{pmatrix}
      1 & T(k)^\dagger\\
      T(k)^\dagger & \mathcal{S}_-
    \end{pmatrix}\Psi_-.
  \end{split}
  \label{eq:action2}
\end{equation}
where the matrix structure is again in the auxiliary index $x$ and we
have introduced the matrices
\begin{equation}
  \mathcal{S}_\pm=T_\pm \mathcal{S} T_\pm\equiv
  \begin{pmatrix}
    T_\pm S T_\pm & 0\\
    0 & T_\pm S^\mathcal{T} T_\pm.
  \end{pmatrix}
\end{equation} 
Here the matrix structure is in time--reversal space and the
time--reversed scattering matrix $S^\mathcal{T}$ has been defined in
\eqref{eq:timerev1}.

\subsection{The color-flavor transformation} 

We are now in a position to subject the generating functional to the
spectral average, $\langle \zeta \rangle_k =\langle \zeta
\rangle_\phi$. As discussed in section \ref{sec:phaseaverage}, we
replace $T(k)\rightarrow T(\{\phi_b\})$, whereupon the average is
given by
\begin{equation}
  \begin{split}
    \langle \zeta(j_+,j_-;s)\rangle_\phi=& \int d(\tilde\psi,\psi)
    \,\mathrm{sdet}(T_+ T_-^\dagger)^2
    \times\\
    &e^{-\bm{S}_0} \prod_{b=1}^B \int \frac{d\phi_b}{2\pi}
    e^{-\bm{S}_{1,b}}.
  \end{split}
  \label{eq:action_phaseaverage}
\end{equation}
Here,
\begin{equation}
  \begin{split}
    \bm{S}_0=&\tilde\Psi_{+,1}\Psi_{+,1}+ \tilde\Psi_{-,1}\Psi_{-,1}
    +\\
    &\tilde\Psi_{+,2}\mathcal{S}_+^\dagger\Psi_{+,2}
    +\tilde\Psi_{-,2}\mathcal{S}_-\Psi_{-,2}
  \end{split}
\end{equation}
is the phase--independent part of the action and
\begin{equation}
  \bm{S}_{1,b}=2\tilde\Psi_{+,1,b}e^{i\phi_b}\Psi_{+,2,b}
    +2\tilde\Psi_{-,2,b}e^{-i\phi_b}\Psi_{-,1,b}.
\end{equation}
So far, we have not achieved much other than representing the spectral
determinants by a complicated Gaussian integral, averaged over phase
degrees of freedom. The most important step in our analysis will now
be to subject the generating function to an integral transform known
as the color--flavor transformation~\cite{Zirnbauer}. The
color--flavor transformation amounts to a replacement of the
phase--integral by an integral over a new degree of freedom, $Z$.
Much better than the original degrees of freedom, the $Z$--field will
be suited to describe the low energy physics of the system.

In a variant adopted to the present context (a single `color' and $F$
`flavors') the color--flavor transformation assumes the form
\begin{multline}
  \int \frac{d\phi}{2\pi} e^{ \eta_{+}^Te^{i\phi}\nu_{+} +
    \nu_{-}^Te^{-i\phi}\eta_{-} }
  =\\
  \int d(\tilde{Z},Z) \mathrm{sdet}\left(1-Z\tilde{Z}\right) e^{
    \eta_{+}^T Z\eta_{-}+ \nu_{-}^T \tilde{Z}\nu_{+} },
  \label{eq:CF_trafo}
\end{multline}
where $\eta_\pm$ and $\nu_\pm$ are arbitrary $2F$ dimensional
supervectors and $Z$, $\tilde{Z}$ are $2F$-dimensional supermatrices.
The boson--boson and fermion--fermion block of these supermatrices are
related by $\tilde{Z}_{\mathbf{bb}}=Z_\mathbf{bb}^\dagger$,
$\tilde{Z}_{\mathbf{ff}}=-Z_\mathbf{ff}^\dagger$, while the entries of
the fermion--boson and boson--fermion blocks are independent
anti--commuting integration variables. The integration
$d(\tilde{Z},Z)$ runs over all independent matrix elements of $Z$ and
$\tilde{Z}$ such that all eigenvalues of
$Z_\mathbf{bb}Z_\mathbf{bb}^\dagger$ are less than unity and the
measure is normalized such that
\begin{equation}
  \int d(\tilde{Z},Z)\, \mathrm{sdet}(1-Z\tilde{Z})=1.
\end{equation}

\begin{figure}[hbt]
  \centerline{\resizebox{6cm}{!}{\includegraphics{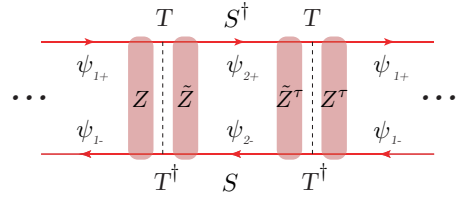}}}
  \caption{On the physical interpretation of the color-flavor
    transformation. Explanation, see text.}
  \label{Zmode}
\end{figure}

We apply the color-flavor transformation $B$ times -- once for each
phase $\phi_b$. As a result, we obtain a $B$--fold integral over
supermatrices $Z_b$. There are four flavors (direction index $d=\pm 1$
and time-reversal index $t=1,2$). We combine all matrices $Z_b$
($\tilde{Z}_b$) into a single block--diagonal $8B$-dimensional
supermatrix $Z$ ($\tilde{Z}$) such that
\begin{equation}
  Z_{bdts,b'd't's'}=\delta_{b,b'} Z_{b\, dts,d't's'}.
\end{equation} 
The averaged generating function now has the form
\begin{equation}
  \begin{split}
    \langle \zeta(j_+,j_-;s) \rangle=& \mathrm{sdet}(T_+T_-^\dagger)^2
    \int d(\tilde\psi,\psi) \int d(\tilde{Z},Z)\\
    & \mathrm{sdet}(1-\tilde{Z}Z)\,
    e^{-\bm{S}(\tilde{\Psi},\Psi,\tilde{Z},Z)}
  \end{split}
\end{equation}
where
\begin{equation}
  \begin{split}
    \bm{S}(\tilde{\Psi},\Psi,\tilde{Z},Z)=& \tilde{\Psi}_1
    \begin{pmatrix}
      \mathbbm{1} & Z\\
      Z^\tau & \mathbbm{1}
    \end{pmatrix}
    \Psi_1+\\
    & \tilde\Psi_2
    \begin{pmatrix}
      \mathcal{S}_+^\dagger & \tilde{Z}^\tau\\
      \tilde{Z} & \mathcal{S}_-
    \end{pmatrix}
    \Psi_2,
  \end{split}
\end{equation}
and we used $2\tilde\Psi_{1}Z\Psi_1=\tilde\Psi_{1}Z\Psi_1+
\tilde\Psi_{1}Z^\tau\Psi_1$,
$2\tilde\Psi_{2}\tilde{Z}\Psi_2=\tilde\Psi_{2}\tilde{Z} \Psi_2+
\tilde\Psi_{2}\tilde{Z}^\tau\Psi_2$. Here, the indices $1$, $2$ refer
to the auxiliary index $x$, and the matrix structure is in
advanced/retarded space.  Integrating the Gaussian fields $\tilde\Psi$
and $\Psi$ we arrive at the (exact) representation
\begin{equation} 
  \langle \zeta(j_+,j_-;s) \rangle=
  \int d(\tilde{Z},Z) e^{-\bm{S}(\tilde{Z},Z)}
\end{equation}
where the action is given by
\begin{equation}
  \begin{split}
    \bm{S}(\tilde{Z},Z)=& -\mathrm{str}\,\mathrm{log}\, (1-\tilde{Z}Z)
    +\frac{1}{2} \mathrm{str}\,\mathrm{log}\,(1- Z^\tau Z) \\&
    +\frac{1}{2} \mathrm{str}\,\mathrm{log}\,(1- \mathcal{S}_+
    \tilde{Z}^\tau \mathcal{S}_-^\dagger \tilde{Z}).
  \end{split}
  \label{eq:exact_action}
\end{equation}
(Note, that the prefactor $\mathrm{sdet}(T_+T_-^\dagger)^2$ has
canceled out.)

Before carrying on, let us pause to discuss the advantage gained by
switching to the $Z$--representation.  Consider the bilinears
$\tilde\psi_{(+/-),s,x,d,b}e^{(+/-)i\phi_b}\psi_{(+/-),s,x,d,b}$
appearing as building blocks of the original phase--representation.
Loosely identifying $\psi_{+/-}$ as retarded/advanced wave function
amplitudes, these products describe the scattering of single particle
states off phase fluctuations. Due to the effective randomness of the
phases they fluctuate in a wild and non--controllable manner (see
Fig.~\ref{Zmode} for a cartoon of the propagation of a retarded [upper
line] and advanced [lower line] wave function in space: a rapid
succession of scattering events [the vertical dashed lines] leads to
strong fluctuations.)  Technically, this means that the original
representation defies controlled evaluation schemes (such as mean
field approximations and the like.)

  In contradistinction, the
$Z$--field enters the theory as $\sim \tilde
\Psi_{+,s,t,d,b}Z_{b,ss',tt',dd'}\Psi_{-,s',t',d',b}$, i.e. through
structures that {\it couple} retarded and advanced amplitudes locally
in space. While (prior to the phase averaging) each of the
$\Psi_{+/-}$ amplitudes individually was a rapidly fluctuating
contribution, the product $\tilde\Psi_+ \Psi_-$ contains benign, slowly
fluctuating contributions. This is because the phase $\exp(i\phi)$
picked up by the retarded amplitude may cancel against the phase
$\exp(-i\phi)$ carried by the advanced amplitude. In a semiclassical
manner of speaking, this happens if the two amplitudes propagate along
Feynman paths locally correlated in space. The advantage of the
$Z$--representation is that it selects precisely these slowly
fluctuating, spatially correlated bilinears which survive the
averaging over phases. In Fig. \ref{Zmode}, the $Z$--fields are
indicated by vertical ovals. Wave function amplitudes qualifying to
form a slowly fluctuating couple may carry different time--reversal
and directional indices which explains the {\it matrix}--structure of
$Z$ in these index spaces.  At any rate, the structure of the
color--flavor transformed theory indicates that the $Z$--integral will
be comparatively benign and amenable to stationary phase treatment.

\section{Saddle point analysis and universality  
\label{sec:saddlepoint}} 

The action \eqref{eq:action2} provides for an exact representation of
the generating functional of an individual graph. While the integral
over $Z$ cannot be done in closed form, it turns out to be ideally
suited to a mean field treatment. In the following, we will formulate
the mean field analysis and explore under which conditions the theory
reduces to one that predicts universal GOE statistics. (We assume time
reversal invariance throughout.)

Our strategy will be to first identify uniform 'zero mode' solutions
to the mean field equations, and the corresponding mean field action.
We will find that the integral over the reduced action generates an
exact RMT expression for the spectral determinants. In a second step
we proceed to investigate the validity of the zero mode approximation,
i.e. we will explore under which conditions corrections to the RMT
result vanish in the semiclassical limit $B\to \infty$.

\subsection{The zero-mode and universality} 

We begin by expanding the full action to linear order in the sources
$p\pm=\Delta(\pm s -\sigma_3^\mathbf{BF} j_\pm)$
\begin{equation}
  \begin{split}
    \bm{S}(\tilde{Z},Z)=&
    \bm{S}_0(\tilde{Z},Z)\\
    &- \frac{i}{4}\mathrm{str} \frac{ p_+(\mathcal{L}\mathcal{S}+
      \mathcal{S}\mathcal{L})\tilde{Z}^\tau \mathcal{S}^\dagger
      \tilde{Z}}{1-\mathcal{S}
      \tilde{Z}^\tau\mathcal{S}^\dagger \tilde{Z}}\\
    &+\frac{i}{4}\mathrm{str} \frac{p_-(\mathcal{L}
      \mathcal{S}^\dagger+ \mathcal{S}^\dagger \mathcal{L})\tilde{Z}
      \mathcal{S}\tilde{Z}^\tau}{1-\mathcal{S}^\dagger
      \tilde{Z}\mathcal{S} \tilde{Z}^\tau}\\
    &+\mathcal{O}\left( (p\mathcal{L})^2\right)
  \end{split}
  \label{eq:exact_action1}
\end{equation}
Here, $\bm{S}_0(Z,\tilde{Z})$ is obtained from \eqref{eq:exact_action}
by replacing $\mathcal{S}_\pm \rightarrow \mathcal{S}$, and
$\mathcal{L}_{bb'}=\delta_{bb'} L_b$ contains the bond length on its
diagonal.
Since we are only interested in spectral fluctuations on the scale of
the mean level spacing $p_\pm \sim \mathcal{O}(B^{-1})$ implying that
higher orders in the expansion in $p_\pm$ vanish in the limit
$B\rightarrow \infty$. At this point we have to assume moderate bond
length fluctuations such that $L_b/\overline{L}\ll B$.

To identify the mean field configurations of the theory, we
differentiate the action $\bm{S}_0$ w.r.t.  $Z$ and obtain
\begin{equation}
  \frac{1}{1-\tilde{Z}Z}\tilde{Z}
  -
  \frac{1}{1-Z^\tau Z} Z^\tau=0
  \label{eq:saddle_eq1}
\end{equation}
This equation is solved by
\begin{equation}
  Z=\tilde{Z}^\tau.
  \label{eq:saddle_sol1}
\end{equation}
Differentiating w.r.t. $\tilde{Z}$ and using \eqref{eq:saddle_sol1} a
second saddle--point equation assumes the form
\begin{equation}
  \begin{split}
    \frac{1}{1-Z\tilde{Z}}Z -\frac{1}{2} \frac{1}{1-\mathcal{S}
      Z\mathcal{S}^\dagger \tilde{Z}}
    \mathcal{S} Z \mathcal{S}^\dagger-&\\
    \frac{1}{2} \frac{1}{1-\mathcal{S}^\tau Z \mathcal{S}^{\dagger
        \tau} \tilde{Z}} \mathcal{S}^\tau Z \mathcal{S}^{\dagger
      \tau}&=0.
  \end{split}
  \label{eq:saddle_eq2}
\end{equation}
This equation is solved by all field configurations that commute with
the scattering operators, i.e.
\begin{equation}
  \begin{split}
    Z_{0b,dts,d't's'}=&\delta_{dd'}
    Y_{ts,t's'}\\
    \tilde{Z}_{0b,dts,d't's'}=& \delta_{dd'} \tilde{Y}_{ts,t's'},
  \end{split}
\end{equation}
which corresponds to equidistribution on the set of directed bonds.
The symmetry condition $\tilde{Z}=Z^\tau$ obtained from the first
saddle--point equation implies $\tilde{Y}=Y^\tau$ where $Y^\tau=\tau_0
Y^T \tau_0^{-1}$ and the matrix $\tau_0$ has been defined
in~\eqref{tau_def}.
The commuting parts of these matrices obey $\tilde{Y}_{\mathbf{bb}}
=Y^*_{\mathbf{bb}}$ and $\tilde{Y}_{\mathbf{ff}} =-Y^*_{\mathbf{ff}}$
while the non--commuting entries are all independent integration
variables.  The fermion-fermion part is integrated over
$\mathbbm{R}^4\simeq \mathbbm{C}^2$ while boson-boson part is
restricted to the compact region where all eigenvalues of
$Y_\mathbf{bb}^\dagger Y_\mathbf {bb}$ are less than unity.

Reducing the action \eqref{eq:exact_action1} to the zero-mode the
first contribution vanishes exactly $\bm{S}_0(Z_0,\tilde{Z}_0)=0$
while the remaining term becomes
\begin{equation}
  \bm{S}^\mathrm{GOE}
  (\tilde{Y},Y)=
  +i\frac{\pi}{\Delta}  
  \mathrm{str}  
  \frac{p_+ 
    Y\tilde{Y}}{1- Y\tilde{Y}}
  -i \frac{\pi}{\Delta}   
  \mathrm{str}  
  \frac{p_- \tilde{Y}Y}{ 
    1-\tilde{Y}Y}, 
  \label{eq:action_GOE}
\end{equation}
Restricting the integration to the zero mode sector, we obtain
\begin{equation}
  \langle \zeta(j_+,j_- ; s)\rangle\simeq   
  \zeta^\mathrm{GOE}(j_+,j_- ; s)\rangle
  \equiv \int d(\tilde{Y},Y) e^{-\bm{S}^\mathrm{GOE}(Y,\tilde{Y})},
\end{equation}
where the denotation $\zeta^\mathrm{GOE}$ indicates that the matrix
integral over $Y$ obtains but an exact representation of the GOE
correlation function. To represent the integral on the r.h.s. in a
more widely recognizable form, let us define the $8 \times 8$
supermatrix
\begin{equation}
  \begin{split}
    Q=&
    \begin{pmatrix}
      \mathbbm{1} & Y\\
      \tilde{Y} & \mathbbm{1}
    \end{pmatrix}
    \Sigma_z
    \begin{pmatrix}
      \mathbbm{1} & Y\\
      \tilde{Y} & \mathbbm{1}
    \end{pmatrix}^{-1}\\
    =&
    \begin{pmatrix}
      1+2 Y\tilde{Y}/(1-Y\tilde{Y}) &
      -2 Y/(1-\tilde{Y}Y)\\
      2 \tilde{Y}/(1-Y\tilde{Y})& -1 -2\tilde{Y}Y/(1-\tilde{Y}Y)
    \end{pmatrix},
  \end{split}
\end{equation}
where $\Sigma_z=\left(
  \begin{smallmatrix} \mathbbm{1}& 0\\
    0 & -\mathbbm{1} \end{smallmatrix}\right)$.  It is then a
straightforward matter to show that the action $\bm{S}(\tilde{Y},Y)$
takes the form of Efetov's action~\cite{Efetov} for the GOE
correlation function
\begin{equation}
\label{Q_int}
   \zeta^\mathrm{GOE}(j_+,j_-;s)=\int dQ\, e^{i \bm{S}(Q)}
\end{equation}
where the measure is given by $dQ\equiv d(\tilde Y,Y)$,
\begin{equation}
  \bm{S}(Q)=\frac{\pi}{2} \mathrm{str}\,
  (Q-\Sigma_z) \hat{\epsilon}
\end{equation}
and $\hat{\epsilon}=-\frac{1}{\Delta}\left(
\begin{smallmatrix}
  p_+ & 0 \\
  0 & p_-
\end{smallmatrix}\right)$. 
For a discussion of the integral \eqref{Q_int}, and the ways
random matrix predictions are obtained by integration over $Q$, we refer to the textbook~\cite{Efetov}.

\subsection{Validity of the saddle--point approximation 
  \label{sec:fluctuations}} 

In the previous section we have shown that the reduction of the theory
to a zero mode integral obtains GOE spectral correlations. However, we
have not yet shown under which conditions this reduction is actually
legitimate. This is the question to which we turn next.

For the purposes of our discussion, it will be sufficient to consider
the expansion of the exact action \eqref{eq:exact_action1} to second
order in the fields $Z$,
\begin{equation}
  \begin{split}
    \bm{S}^{(2)}(Z,\tilde{Z})=&
    \bm{S}_0^{(2)}(Z,\tilde{Z})\\
    &- \frac{i}{4}\mathrm{str} \left(p_+(\mathcal{L}\mathcal{S}+
      \mathcal{S}\mathcal{L})\tilde{Z}^\tau
      \mathcal{S}^\dagger \tilde{Z}\right)\\
    & +\frac{i}{4}\mathrm{str} \left(p_-(\mathcal{L}
      \mathcal{S}^\dagger+ \mathcal{S}^\dagger \mathcal{L})\tilde{Z}
      \mathcal{S}\tilde{Z}^\tau\right),
  \end{split}
\end{equation}
where
\begin{equation}
  \bm{S}_0^{(2)}(Z,\tilde{Z})=
  \mathrm{str}\left(
    \tilde{Z}Z-\frac{1}{2}Z^\tau Z -\frac{1}{2}
    \mathcal{S}_-^\dagger 
    \tilde{Z}\mathcal{S}_+\tilde{Z}^\tau
  \right).
  \label{eq:quadratic_action}
\end{equation}
Physically, the quadratic action describes the joint propagation of a
retarded and an advanced Feynman amplitude along the same path in
configuration space. (This is a generic feature of second order
expansions to nonlinear $\sigma$--models of disordered and chaotic
systems.  For a discussion of this point, we refer to
Ref.~\cite{Efetov}.) It thus carries information similar to that
obtained from the diagonal approximation to semiclassics. The second
order expansion is justified if the fluctuations of the fields $Z$ are
massively damped (in the sense that the matrix elements of $Z$
effectively contributing to the integral are much smaller than unity.)
Under these conditions, the integration over matrix elements of $Z$
may be extended to infinity and we obtain a genuine Gaussian integral.

The eigenvalues of the quadratic form appearing in $\bm{S}^{(2)}$ at
$p_\pm=0$ determine the damping $m_j$ --- or the mass, in a field
theoretical jargon --- inhibiting fluctuations of the eigenmodes
$Z_j$. As indicated by its name, the zero--mode $Z_0$ carries zero
mass. Within the quadratic approximation, the correlation function
assumes the form $\langle \zeta(J_+,j_-;s)\rangle\simeq \langle
\zeta(J_+,j_-;s)\rangle^{(2)} = \prod_j I_j$, with the Gaussian
integrals
\begin{equation}
  \begin{split}
    I_j=\int d(Z_j,\tilde{Z}_j)&
    \exp[-\mathrm{str}\{ m_j Z_j\tilde{Z}_j \\
    & + i (m-1)\overline{L}(p_+Z_j\tilde{Z}_j -p_-\tilde{Z}_jZ_j)\}]
  \end{split}
\end{equation}
where $Z_j$, and $\tilde{Z}_j$ are $4\times 4$ supermatrices obeying
the ubiquitous condition $\tilde{Z}_{j\mathbf{bb}}=Z^*_{j\mathbf{bb}}$
and $\tilde{Z}_{j\mathbf{ff}}=-Z^*_{j\mathbf{ff}}$. We also assumed
here, that the first saddle--point equation $Z=\tilde{Z}^\tau$ is
obeyed which reduces the number of integration variables by a factor
$1/2$.  Configurations which are orthogonal to this condition give the
same kind of factors but have different masses.  Doing the Gaussian
integral~\cite{Gaussian} we obtain
\begin{equation}
  I_m=\frac{[1+i\frac{\pi(m-1)(s+j_\Sigma)}{mB}]^2
    [1+i\frac{\pi(m-1)(s-j_\Sigma)}{mB}]^2
  }{[1+i\frac{\pi(m-1)(s+j_\Delta)}{mB}]^2
    [1+i\frac{\pi(m-1)(s-j_\Delta)}{mB}]^2},
\end{equation}
where $j_\Delta=j_+-j_-$ and $j_\Sigma=j_++j_-$. Differentiating
w.r.t. the sources we finally obtain the quadratic approximation to
the correlation function,
\begin{equation}
  \begin{split}
    R^{(2)}_2(s)=&\sum_m \frac{1}{8\pi^2}\mathrm{Re}\,
    \frac{\partial^2}{\partial j_+ \partial j_-}
    I_m \big|_{j_\pm=0}\\
    =&\sum_m \frac{(m-1)^2(m^2 B^2 - \pi^2 (m-1)^2 s^2)}{ (m^2 B^2 +
      \pi^2 (m-1)^2 s^2)^2}.
  \end{split}
\end{equation}
The contribution of the zero mode ($m=0$) is given by $-\frac{1}{\pi^2
  s^2}$ and coincides with the diagonal approximation to the GOE
correlation function. (Later on we shall see that in the case of
broken time reversal invariance, one half of the matrix elements of
$Z_0$ become massive implying that the contribution of the zero mode
reduces to the GUE expression $-\frac{1}{2\pi^2 s^2}$.)

In the limit $B\to \infty$, the $s$-dependence of the contribution of
massive modes to the correlation function is negligible for our
purpose, i.e.  individual modes contribute maximally as
$\sim(m-1)^2/2m^2 B^2\sim (mB)^{-2}$.  Only modes
of mass $m\sim B^{-\alpha}$, where $\alpha$ is a non--vanishing
positive exponent, can survive the limit $B\to \infty$.  The contribution
of an individual mode is negligible if the exponent $0\le \alpha <1$.
There are at most $\mathcal{O}(B)$ nearly massless
modes, and we are led to require that $ B^{2\alpha-1}$ must vanish in
the limit of large graphs $B\rightarrow \infty$, or that $0\le \alpha
<1/2$.

After these general remarks, let us discuss the masses that actually
appear in the quadratic action \eqref{eq:quadratic_action}.  We first
show that modes violating the first saddle--point equation
$Z=\tilde{Z}^\tau$ can safely be neglected.  This is seen by rewriting
the quadratic action as $\bm{S}^{(2)}= \frac{1}{2}\mathrm{str}
\left[(Z-\tilde{Z}^\tau)(\tilde{Z}-Z^\tau)+ \tilde{Z}^\tau \tilde Z -
  \mathcal{S}_+ \tilde{Z}^\tau \mathcal{S}_-^\dagger
  \tilde{Z}\right]$.  This expression shows that fluctuations away
from the condition $Z=\tilde Z^\tau$ are suppressed by a large mass of
$\mathcal{O}(1)$. These fluctuations may safely be ignored, i.e. we
may assume the condition $Z=\tilde Z^\tau$ to be rigidly imposed.
The quadratic action then assumes the reduced form
\begin{equation}
  \bm{S}^{(2)}(Z)=
  \frac{1}{2}\mathrm{str}\left[ Z\tilde Z - 
    \mathcal{S}_+ Z \mathcal{S}_-^\dagger \tilde{Z}\right]
  \label{eq:quadratic_action1}
\end{equation}
where the condition $Z=\tilde{Z}^\tau$ reduces the number of
independent integration variables by a factor one half.

We next show that fluctuations $Z_{d,d'}, d\not=d'$ off--diagonal in
the directional indices may safely be discarded, too. To this end, let us
separate the contribution of diagonal and off--diagonal fields,
$Z^{\rm diag}$ and $Z^{\rm off}$, respectively, to the quadratic
action \eqref{eq:quadratic_action1}:
\begin{equation}
  \mathrm{str}(\mathcal{S}Z\mathcal{S}^\dagger\tilde{Z})=
  \mathrm{str}
  \begin{pmatrix}
    \tilde{Z}^\mathrm{diag}& \tilde{Z}^\mathrm{off}
  \end{pmatrix}
  \begin{pmatrix}
    \mathcal{F} & \mathcal{G} \\
    \mathcal{H} & \mathcal{K}
  \end{pmatrix}
  \begin{pmatrix}
    Z^\mathrm{diag}\\
    Z^\mathrm{off}
  \end{pmatrix}.
\end{equation} 
The matrices $\mathcal{G}$ and $\mathcal{H}$ contain elements of the
type $S_{b\, -d, b'd'}S^*_{bd,b'd'}$ or $S_{b'\, -d', b
  d}S^*_{bd,b'd'}$.  If $S^*_{bd,b'd'}$ does not vanish there must be
a vertex $v$ in the graph, such that the directed bond $(b',d')$ ends
at $v$ and $(b,d)$ starts at $v$. The partner factors $S_{b\, -d,
  b'd'}$ and $S_{b'\, -d', b d}$ then vanish (unless the bond $b$ is a
loop such that $(b,d)$ and $(b,-d)$ both start and end at the vertex
$v$. However, for simple graphs no loops are present and
$\mathcal{G}=\mathcal{H}=0$.)  The matrix $\mathcal{K}$ contains
elements of the form $S_{b\, -d,b'\, -d'}S^*_{bd, b'd'}$ or $S_{b' d',
  b d}S^*_{bd, b'd'}$. For $b\not=b'$, these vanish (unless the bonds
$b$ and $b'$ connect the same pair of vertices which, however, is
forbidden for simple graphs.) For $b=b'$, the non--vanishing of the
matrix element would again require the existence of loops. We thus
conclude that $\mathcal{K}=0$. Decoupled from the scattering
operators, the integration over modes $Z_{d\not=d'}$ merely produces a
factor of unity (supersymmetry!) so that we will concentrate on the
complementary set of modes
\begin{equation}
  \begin{split}
    Z_{b\, dd'}&=\delta_{d d'} Z_{bd}\\
    \tilde{Z}_{b\, dd'}&=\delta_{d d'} \tilde{Z}_{bd},
  \end{split}
\end{equation}
throughout.  The contribution of these configurations to the
generating function is determined by the elements of the matrix
$\mathcal{F}=\{|S_{bd,b'd'}|^2\}$. (Here, we used that for a time
reversal invariant graph, $|S_{bd,b'd'}|^2 =S_{bd,b'd'} S_{b'\,
  -d',b\, -d}^*$, i.e. that the matrix $\mathcal{F}$ is isotropic in
time reversal space.) Specifically, the action ${\bf S}_0$ assumes the
form
\begin{equation}
   \bm{S}^{(2)}={1\over 2}\mathrm{str}(
   \tilde{Z}(\mathbbm{1}-\mathcal{F})Z)
  \label{eq:second_order_action3}
\end{equation}
Within the context of the semiclassical analysis of appendix
\ref{sec:tanner}, we have seen that the matrix $\mathcal{F}$
determines the classical propagator (the Frobenius--Perron operator)
on the graph. Comparing with our discussion above, we conclude that
the eigenvalues of that operator, $\lambda_i$, determine the `mass
spectrum' $\{m_i = 1-\lambda_i\}$ of the theory. We have seen that
large graphs behave universal if the masses scale as $m_i\sim
B^{-\alpha}$, $0\le \alpha<1/2$. Specifically, this condition requires
the gap $1-\lambda_1$ between the zeroth Perron--Frobenius eigenvalue
$\lambda_0=1$ (corresponding to the fully equilibrated zero--mode
configuration) and the first `excited' state to scale as
$1-\lambda_1\equiv \Delta_g\gtrsim B^{-\alpha}, 0\le \alpha<1/2$. This
condition is stricter then Tanners conjecture
$0\le \alpha<1$: For $\alpha\ge 1$, corrections to the universal
result remain sizeable no matter how large the graph is. In the
intermediate region $1/2 \le \alpha <1$ --- permissible by Tanner's
criterion --- non--universal corrections vanish only if the number $r$
of classical modes with a small mass remains constant (or does not
grow too fast) such that $B^2 \Delta_g^2/r \rightarrow \infty$. If,
however, the number of low energy modes is extensive, $r\sim B$, the
stricter condition $0\le \alpha<1/2$ has to be imposed to stabilize
universality.

Above we have shown that in the limit $B\to \infty$ only the zero mode
effectively contributes to the correlation function (provided, of
course, the master condition $\Delta_g \sim B^{-\alpha}$ is met.)
While the zero mode integral must be performed rigorously, all other
modes are strongly overdamped and may be treated in a quadratic
approximation. (This is the a posteriori justification for the
quadratic approximation on which our analysis of the mass spectrum was
based.)

\subsection{GOE-GUE crossover 
  \label{sec:crossover}} 

The analysis above applied to time reversal invariant graphs. In this
section we discuss what happens if time reversal invariance gets
gradually broken, e.g. by application of an external magnetic field.  
We assume full universality, i.e. $B \Delta_g^2 \gg 1$ such that only
the zero--mode contributes to $R_2(s)$. Our aim is to derive a
condition for the crossover between GOE--statistics (time reversal
invariance) and GUE--statistics (lack of time reversal invariance.)

The substructure of the $Z$--fields in time reversal space is given by
\begin{equation}
    Z_b=
    \begin{pmatrix}
      Z_{\mathbf{D},b}& Z_\mathbf{C,b}\\
      \tilde{Z}_{\mathbf{C},b}^T\sigma_3^\mathbf{BF} &
      \tilde{Z}_{\mathbf{D},b}^T
    \end{pmatrix},\qquad
    \tilde{Z}_b=
    \begin{pmatrix}
      \tilde{Z}_{\mathbf{D},b} & \sigma_3^\mathbf{BF}
      Z_{\mathbf{C},b}^T\\
      \tilde{Z}_{\mathbf{C},b} & Z_{\mathbf{D},b}^T
    \end{pmatrix},
  \label{eq:zmode_goe}
\end{equation}
where $Z_{\mathbf{D/C},b}$ and $\tilde Z_{\mathbf{D/C},b}$ are $2\times 2$
supermatrices subject to the constraint $\tilde{Z}_{\mathbf{bb}}
=Z^*_{\mathbf{bb}}$ and $\tilde{Z}_{\mathbf{ff}} =-Z^*_{\mathbf{ff}}$,
while the non--commuting entries of these matrices are independent
integration variables. The subscripts $\mathbf{D} (\mathbf{C})$ allude
to the fact that in disordered fermion systems, the modes
$Z_\mathbf{D}$ ($Z_\mathbf{C}$) generate the so--called diffuson
(Cooperon) excitations. Physically, the former (latter) describe the
interference of two states as they propagate along the same path (the
same path yet in opposite direction) in configuration space; Cooperon
modes are susceptible to time reversal invariant breaking
perturbations.

Substituting this representation into the quadratic action, we obtain
\begin{equation}
   \bm{S}^{(2)}=\mathrm{str}\left(
   \tilde{Z}_\mathbf{D}(\mathbbm{1}-\mathcal{F}_\mathbf{D})Z_\mathbf{D}
+\tilde{Z}_\mathbf{C}(\mathbbm{1}-\mathcal{F}_\mathbf{C})Z_\mathbf{C}\right)
  \label{eq:second_order_action_CD}
\end{equation}
as a generalization of Eq.~\eqref{eq:second_order_action3}. Here,
$\mathcal{F}_\mathbf{D}=\{|S_{bd,b'd'}|^2\}$ while
$\mathcal{F}_\mathbf{C}= S_{bd,b'd'} S_{b'\, -d',b\, -d}^*$. For a
time reversal non--invariant graph $\mathcal{F}_\mathbf{D}\not=
\mathcal{F}_\mathbf{C}$ and the symmetry of the action in time
reversal invariance space gets lost.

Noting that $2B=\sum_{bd,b'd'} |S_{bd,b'd'}|^2$, we conclude that the
Cooperon zero mode $Z_{\mathbf{C},b}=Y_{\mathbf{C}}$ 
acquires a mass term $\sim B
m_\mathbf{C}\,\mathrm{str}(Y_\mathbf{C}\tilde Y_\mathbf{C})$, where
the coefficient
\begin{equation}
  \begin{split}
    m_\mathbf{C}&=\frac{1}{2B} \left|\sum_{bd,b'd'}
      S_{bd, b'd'}(S^*_{bd, b'd'}-S^*_{b' d'^{-1},b d^{-1}})\right|\\
    & =\frac{1}{B} \left|\mathrm{tr}\,
      S^\dagger(S-\sigma_1^\mathrm{dir} S^T
      \sigma_1^\mathrm{dir})\right|
  \end{split}
\end{equation} 
measures the degree of the breaking of the symmetry $S=S^\mathcal{T}$.
The Cooperon mode may be neglected once $B m_\mathbf{C} \rightarrow
\infty$ as $B\rightarrow \infty$.

For the sake of definiteness, let us discuss two concrete mechanisms
of symmetry breaking: \textit{i.} breaking the time--reversal symmetry
of vertex scattering matrices, and \textit{ii.} application a magnetic
field.

Beginning with \textit{i.}, let us consider a large {\it complete}
graph for simplicity. (A graph is complete if any two of its vertices
are connected by a bond.)  The number of vertices of these graphs is
order $B^{1/2}$, and each column in $S_{b'd',bd}$ has $B^{1/2}$
non--vanishing entries of order $B^{-1/4}$.  Breaking time--reversal
symmetry at a single vertex thus results in a coefficient
$m_\mathbf{C} \sim B^{-1}$.  Breaking time--reversal invariance at a
single vertex is, thus, not sufficient to drive the crossover to GUE
statistics. Rather, a finite fraction $\sim B^{\beta}$ ($\beta>0$) of
time--reversal non--invariant vertices is required. Obtained for the
simple case of complete graphs, it is evident that this conclusion
generalizes to generic graphs.

Turning to \textit{ii.}, the application of a constant magnetic field
$A_b=A$ causes a global change of all its bond scattering matrices; We
have to replace $T(k)=e^{ik \mathcal{L}/2} \rightarrow T(k)
T(\sigma_3^\mathbf{dir} A)= e^{i(k+\sigma_3^\mathbf{dir}A)
  \mathcal{L}/2}$.  This is equivalent to replacing $S\rightarrow
S(A)=T(\sigma_3^\mathbf{dir} A) S T(\sigma_3^\mathbf{dir} A)$ in the
quantum map.  Assuming time--reversal invariance at $A=0$ the mass of
the cooperon mode becomes
\begin{equation}
  \begin{split}
    m_\mathbf{C}(A)&=\frac{1}{2B} \left|\mathrm{tr}\, S^\dagger [S -
      T(-\sigma_3^\mathbf{dir}A)S
      T(-\sigma_3^\mathbf{dir}A)]\right|\\
    &\approx A^2 \mu_2 +\mathcal{O}(A^3)
  \end{split}
\end{equation}
where we used $\sigma_1 \sigma_3 \sigma_1=-\sigma_3$.  We may estimate
\begin{equation}
  \mu_2\approx \frac{\overline{L}}{4B}\left|\mathrm{tr}
  \, \left(S^\dagger \sigma_3^\mathbf{dir} 
    S \sigma_3^\mathbf{dir}  -1 \right)\right|
\end{equation}
by setting $T(-\sigma_3^\mathbf{dir} A)\approx e^{-i\frac{A
    \overline{L}}{2} \sigma_3^\mathbf{dir}}$. For a generic scattering
matrix one may expect $|\mathrm{tr} \, \left(S^\dagger
  \sigma_3^\mathbf{dir} S \sigma_3^\mathbf{dir} -1 \right)|\sim B$
such that a small magnetic field of order $A\sim B^{-1/2}$ is strong
enough to induce the crossover to GUE statistics. (Assuming that the
geometric `area' of the graph, $S\sim B$, is proportional to the
number of bonds, we conclude that the crossover takes place once a
finite number $\sim AS \times e/h \sim B^{0}$ of flux quanta pierces
the system. This crossover criterion is known to apply quite
generically in disordered or chaotic quantum systems.)

\section{Conclusion} 

To summarize, we have shown that the two--point spectral correlation
function of individual quantum graphs coincides with the prediction of
random--matrix theory. Corrections to universality vanish in the limit
$B\rightarrow \infty$ provided the gap in the spectrum of the
underlying `classical' propagator remains constant, or vanishes as
$|m|=|1-\lambda_2| \sim B^{-\alpha}$ with $0\le \alpha < 1/2$. These results
were obtained by representing the generating functional of the
two--point correlation function in terms of a 
nonlinear $\sigma$--model. Closely
resembling the theory of spectral correlations in disordered fermion
systems, this formalism obtained a fairly accurate picture of
correlations in the graph spectrum. Specifically, (i) a perturbative
expansion of the $\sigma$--model for large energies establishes the
contact with semiclassical approaches to the problem, (ii) for low
energies a non--perturbative integration over the fully phase--space
equilibrated zero mode configuration of the model obtains spectral
correlations as predicted by random matrix theory, and (iii) the
analysis of the `mass spectrum' of non--uniform modes yields
conditions under which universality is to be expected: In the limit of
large graph size, $B\to \infty$, the first non--vanishing eigenvalue
of the classical Perron--Frobenius operator on the graph must be
separated from unity by scale $\Delta_g$ larger than $\mathrm{const.}
\times B^{-\alpha}$, $0\le \alpha < 1/2$.

This condition turns out to be met by many prominent classes of
quantum graphs.  Examples include complete DFT graphs, or complete
Neumann graphs~\cite{greg}.  It has also been shown that almost all
unistochastic matrices, i.e. matrices of the type
$\mathcal{F}_{ij}=|S_{ij}|^2$, where $S$ runs over the unitary group
(or, equivalently the circular unitary ensemble CUE), have a finite
gap in the limit of large matrices \cite{gregongap}.  For example,
star graphs~\cite{stars} with the central vertex scattering matrix $S_{ij}$
generically display universal spectral statistics. (Counterexamples
such as the Neumann star graph~\cite{gregonstars}  are not generic in this sense.)

Cases where the universality condition is violated include large
graphs with low valency (coordination number) of vertices. In such
systems Anderson localization phenomena may interfere with the buildup
of universal correlations\cite{schanz,solomyak}.

\appendix

\section{Periodic orbit theory for graphs and  
  Tanner's conjecture \label{sec:tanner}}
 
For completeness we briefly review some elements of the periodic orbit
approach to spectral statistics on quantum graphs. Central to the
periodic orbit approach is a short-time expansion of the spectral form
factor, $K(\tau)$, the Fourier transform of the two--point correlation
function $R_2(s)=\int_0^\infty K(\tau) e^{i2 \pi s \tau}$.  For
moderate bond length fluctuations in quantum graphs this quantity is
usually replaced by the essentially equivalent quantity
\begin{equation}
  K_n=\frac{1}{2 B}\left\langle 
  \mathrm{tr}\, \mathcal{U}_B(k)^n
  \mathrm{tr}\, {\mathcal{U}_B(k)^\dagger}^n
  \right\rangle_k.
\label{eq:discrete_ff}
\end{equation}
Here $n\in \mathbbm{N}$ is a discrete time corresponding to
$\tau={n}{2B}$.  This discrete version of the form factor is connected
to the correlations in the eigenphases of the quantum map.

The form factor \eqref{eq:discrete_ff} is a double sum over periodic
orbits $\gamma,\gamma'$ of lengths $n$ where a periodic orbit on the
graph is a periodic sequence of directed bonds visited.  In the
diagonal approximation --- valid for short times $n\ll 2B$ --- only
those pairs of orbits are taken into account where $\gamma'=\gamma$,
or $\gamma'=\gamma^T$ where $\gamma^T$ is the time-reversed periodic
orbit:
\begin{equation}
  K_n^\mathrm{diag}= \frac{n}{B} \mathrm{tr}\, F^n
  \label{eq:formfactor_diagonal}
\end{equation}
where
\begin{equation}
  F_{bd,b'd'}=|S_{bd,b'd'}|^2
  \label{eq:unistochastic}
\end{equation}
is the 'classical' propability to be scattered from the directed bond
$(b',d')$ to $(b,d)$.  It can be considered as the equivalent of the
Frobenius--Perron propagator for Hamiltonian flows.  Due to unitarity
of $S$ the matrix $F$ is \emph{bistochastic} $\sum_{bd} F_{bd,b'd'}=
\sum_{bd} F_{b'd',bd}=1$ and describes a Markov process on the
directed bonds of the graph.
The eigenvalues $\lambda_i$ of bistochastic matrices are known to lie
in the unit circle $|\lambda_i|\le 1$ with at least one eigenvalue
unity (here corresponding to equidistribution on the directed bonds.)

Universal spectral statistics is expected for `chaotic' graphs in the
limit $B\rightarrow \infty$. According to
Eq.~\eqref{eq:formfactor_diagonal}, the necessary condition for
universality is given by
\begin{equation}
  \mathrm{ tr}\, F^{\tau 2B}\rightarrow 1,
  \label{eq:condition1}
\end{equation}
where the scaled time $\tau\equiv n/2B\ll 1$ is kept constant and the
limit $B\to \infty$ is implied. In this case, $K^{\rm diag} \to 2\tau$
in agreement with the short time expansion
$K^\mathrm{GOE}(\tau)\stackrel{\tau\ll 1}{\simeq} 2\tau$ of the
RMT--form factor
$$
K^\mathrm{GOE}(\tau)=
\begin{cases}
  |\tau|\left(2 -\log(2|\tau|+1)\right)&
  \text{for $|\tau| < 1$}\\
  2-|\tau| \log\frac{2|\tau|+1}{2|\tau|-1} & \text{for $|\tau| \ge
    1$}.
\end{cases}
$$
Here, $\tau\equiv t \Delta/2\pi$ is time measured in units of the
RMT--level spacing.

The universality condition above states that any propability
distribution on the graph will eventually decay to
equidistribution -- a Markov process with this property
is called `mixing' which implies ergodicity (equality of long time-averages 
to an an average over the equidistribution on bonds). 
This is very week condition on a connected
graph: a non-ergodic Markov map on a graph implies equivalence to
a Markov map on a disconnected graph.  However, as observed by
Tanner~\cite{Tanner} the condition~\eqref{eq:condition1} is
actually stronger than mixing dynamics (for an example
of a mixing graph with non--universal spectral statistics 
--- the Neumann star graph, see~\cite{gregonstars}).  This can be
seen by rewriting the~\eqref{eq:condition1} in terms of the $2B$
eigenvalues $\lambda_i$ of $F$.  Ordering the eigenvalues in
magnitude such that $1=\lambda_1\ge|\lambda_2|\ge\dots$ and
defining the spectral gap
\begin{equation}
  \Delta_\mathrm{gap} =1-|\lambda_2|,
\end{equation}   
mixing dynamics merely implies $\Delta_\mathrm{gap}>0$, i.e. that
$\lambda_1=1$ is the only eigenvalue on the unit circle. However,
there are examples of graphs 
whose `classical' dynamics is mixing while the
form factor does not start as $2\tau$. To understand the origin of
this exceptional behavior, notice that
\begin{equation}
  \mathrm{tr}\, F^{n}=\sum_{i=1}^{2B} 
  \lambda_i^{n}=1 + \sum_{i=2}^{2B} \lambda_i^{n},
\end{equation} 
implying the universality criterion $|\sum_{i=2}^{2B}
\lambda_i^{n}|\rightarrow 0$.  With $|\sum_{i=2}^{2B}
\lambda_i^{2B\tau}|\le \sum_{i=2}^{2B} |\lambda_i|^{2B\tau}\le (2B-1)
(1-\Delta_\mathrm{gap})^{2 B\tau}$.  If $\Delta_\mathrm{gap}$ remains
a finite constant as $B\rightarrow \infty$ this surely vanishes and
universality is guaranteed.  However if $\Delta_\mathrm{gap}= c
B^{-\alpha}$ the correction to unity vanished only if $0\le \alpha <
1$.  In all known examples of graphs where the spectral statistics is
non--universal in spite of ergodic classical dynamics this condition
is, indeed, violated: $\alpha\ge 1$. For this reason Tanner
conjectured that $0\le \alpha < 1$ is a sufficient universality
criterion {\it in the scaling limit} $\tau$ fixed while $B\rightarrow
\infty$.

\section{Time--reversal invariant graphs with spin}
\label{sec:GSE}

In this appendix we discuss the spectral statistics of time--reversal
invariant graphs with spin (symmetry class $A$II, or symplectic
symmetry.) This case has been considered in connection with the Dirac
equation on graphs~\cite{Harrison}.  Following a somewhat different
approach, we will here break spin rotational invariance by choosing
vertex boundary conditions that couple different spin components (yet
leave time--reversal invariance intact.)

A spin degree of freedom is straightforwardly introduced by adding a
spin component $m =\pm \frac{1}{2}$ to the wave function on the bonds.
This extension turns the quantum evolution map $\mathcal{U}_B(k)=T(k)
S T(k)$ into a $4B \times 4B$ matrix. We consider non--magnetic graphs
($A_b=0$) and assume independent propagation of the spin components on
the bonds: $T(k)_{bdm,b'd'm'}=\delta_{b,b'} \delta_{d,d'}
\delta_{m,m'} e^{ik L_b/2}$.  Mixing of spins occurs at the vertex
scattering centers. Comprising all vertex scattering matrices into a
single unitary matrix $S$ (defined in analogy to spinless case
discussed in the text) we obtain the condition
\begin{equation}
  S=S^\mathcal{T}\equiv 
  \sigma^\mathbf{dir}_\mathrm{1}
  \sigma^\mathbf{spin}_\mathrm{2}
  S^T \sigma^\mathbf{spin}_\mathrm{2}
  \sigma^\mathbf{dir}_\mathrm{1}
  \label{eq:timerev2}
\end{equation}
for time-reversal invariance
where $\sigma^\mathbf{spin}_\mathrm{2} =i\left(
  \begin{smallmatrix}
    0 & -\mathbbm{1}\\
    \mathbbm{1} & 0
  \end{smallmatrix}
\right)$ is the Pauli matrix in spin indices $m$.  The corresponding
anti--unitary time--reversal operator obeys
$\mathcal{T}^2=-\mathbbm{1}$ characteristic of symmetry class is
$A$II.

Any system in class $A$II has a doubly degenerate spectrum due to
Kramers' degeneracy. By convention, each of the doubly degenerate
eigenvalues is counted only once, i.e.  the mean level spacing
$\Delta=\pi/B\overline{L}$ as with spinless graphs. The oscillatory
contribution to the density of states is given by $ \delta d(k) = -
\frac{1}{2 \pi} \mathrm{Im}\, \frac{d}{dk} \mathrm{ln}\, \xi(k^+).  $

In essence, the derivation of field theory representation of the
generating function parallels the spinless case. However, the
generalized transposition is now defined as
\begin{equation}
  \begin{split}
    \tau=&- i\sigma_1^\mathbf{dir}\sigma_2^ \mathbf{spin} \tau_0
    \\
    \tau_0=& -i E_\mathbf{b}\sigma_2^\mathbf{tr} +
    E_\mathbf{f}\sigma_1^\mathbf{tr},
  \end{split}
\end{equation}
such that $\tau^T=\tau^{-1}$ and $\tau^2=-\tau_0^2=
\sigma_3^\mathbf{BF}$.  The definition of $\tau_0$ reflects the
different type of time--reversal symmetry.  Second, the presence of a
spin component implies that the matrices $Z_b$ and $\tilde{Z}_b$ ---
introduced by color flavor transformation as before --- now have
dimension $16 \times 16$.  These differences understood,
\eqref{eq:exact_action} applies to graphs with spin.

The saddle--point conditions identify a zero--mode diagonal in
directional, and in spin indices $Z_{0b\,dmts,d'm't's'}=\delta_{dd'}
\delta_{mm'}Y_{ts,t's'}$ ($\tilde{Z}_{0b\,dmts,d'm't's'}=\delta_{dd'}
\delta_{mm'}Y_{ts,t's'}$) with $\tilde{Y}=Y^\tau$ where $Y^\tau=\tau_0
Y^T \tau_0^{-1}$.  In an explicit way of writing, the time reversal
structure of the $Z$--matrices (or, equivalently, the $Y$--matrices)
is given by
\begin{equation}
  \begin{split}
    Z=&
    \begin{pmatrix}
      Z_{\mathbf{D}}& Z_\mathbf{C}\\
      -\tilde{Z}_\mathbf{C}^T\sigma_3^\mathbf{BF} &
      \tilde{Z}_\mathbf{D}^T
    \end{pmatrix}\\
    \tilde{Z}=&
    \begin{pmatrix}
      \tilde{Z}_\mathbf{D} &
      -\sigma_3^\mathbf{BF} Z_\mathbf{C}^T \\
      \tilde{Z}_\mathbf{C} & Z_\mathbf{D}^T
    \end{pmatrix},
  \end{split}
  \label{eq:zmode_gse}
\end{equation}
Projected onto the zero--mode sector, the action is given by
\begin{equation}
  \bm{S}^\mathrm{GSE}(\tilde{Y},Y)=
  +i\frac{2 \pi}{\Delta} 
  \mathrm{str} 
  \frac{p_+
    Y\tilde{Y}}{1- Y\tilde{Y}}
  -i \frac{2 \pi}{\Delta}  
  \mathrm{str} 
  \frac{p_- \tilde{Y}Y}{
    1-\tilde{Y}Y}.
  \label{eq:action_GSE}
\end{equation}
Except for a factor $2$ (which accounts for Kramers' degeneracy) this
expression equals the GOE action \eqref{eq:action_GOE}; the difference
between the two cases is hidden in the symmetry $Y=\tilde{Y}^\tau$.
As with the GOE case above, an integration over the matrices
$Y$~\cite{Efetov} obtains the correlation function of the GSE.

Sufficient conditions for the zero--mode reducibility can be derived
as in the spinless case.  Modes violating the symmetry condition
$Z=\tilde{Z}^\tau$, and modes off-diagonal in direction space may be
discarded.  Similarly, modes off--diagonal in the spin index $m$ are
massive and may be neglected, too. Therefore, the validity of the
saddle--point reduction again relies on the discreteness of the
eigenmode spectrum of the `classical' propagator $\mathcal{F}_{bdm,
  b'd'm'}=|S_{bdm,b'd'm'}|^2$.

\section{Color--flavor transformation}
\label{sec:color-flav-transf}

In this appendix we sketch the main conceptual input entering the
proof of the color--flavor transformation (\ref{eq:CF_trafo}). (For a
detailed exposure of the proof, we refer to \cite{Zirnbauer}.) Central
to our discussion will be a (super)--algebra of operators defined as
\begin{equation}
  \label{cf:super_alg} 
  [c_A,c_A']\equiv c_A \bar c_A' -
  (-)^{|A||A'|} \bar c_{A'}
  c_{A}=\delta_{A,A},  
\end{equation} 
where $A=(a,\alpha)$, the index $\alpha,\alpha'$ keeps track of all
flavor components of the theory (boson/fermion, time--reversal,
directional, etc.), and $a,a'=\pm$ distinguish between retarded and
advanced indices. These operators act in an auxiliary Fock--space,
whose vacuum state is defined by the condition
$$
c_{+,\alpha}|0\rangle = \bar c_{-,\alpha}|0\rangle = 0,\qquad
\langle 0 | \bar c_{+,\alpha} = \langle 0 | c_{-,\alpha}=0.
$$
We may, thus, think of the vacuum as a configuration where all
$(+)$--states are empty, while all $(-)$--states are filled;
Excitations are formed by creating $+$--particles and $(-)$--holes.

The above Fock space contains a sub--space defined by the condition
that each state contains as many excited $(+)$--particles as
$(-)$--holes.  We call this space as the flavor--space.
(Alternatively, we may characterize the flavor--space as the space of
all excitations that can be reached from the vacuum state without
changing the number of particles.)  In essence, color--flavor
transformation amounts to the construction of two different
representations of a projector onto the flavor space.

Representation no.~1 is constructed as follows: a generic state in
Fock space can be represented as a linear combination $|\Psi \rangle =
\sum_{n_+,n_-} |\Psi_{n_+,n_-}\rangle$, where $|\Psi_{n_+,n_-}\rangle$
contains $n_{+/-}$ particle/hole states. A projection onto the flavor
state can now be trivially effected by mapping
$$
|\Psi\rangle \mapsto {\cal P}^I |\Psi\rangle \equiv {1\over 2\pi}
\int d\phi\, e^{i(n_+-n_-) \phi} |\Psi_{n_+,n_-}\rangle.
$$

To construct representation no.~2 some more of preparatory work is
necessary: Consider the Lie--supergroup ${\rm Gl}(2F,2F)$ (the group
of $(4F)$--dimensional invertible supermatrices.) This group acts in
Fock space by the representation ${\rm Gl}(2F,2F)\ni g \mapsto
T_g\equiv \exp(\bar c_A \, \ln(g)_{A,A'}\, c_{A'})$, where $T_g$ acts
as linear map in Fock space. (For the proof that the assignment
$g\mapsto T_g$ meets all criteria required of a group representation,
see \cite{Zirnbauer}.)

One may show that the representation above is irreducible in flavor
space. This implies that the entire space may be generated by the
action of the group on a fixed reference state, the vacuum state, say.
The set of `coherent states' $|g\rangle \equiv T_g |\Omega\rangle$
will play a crucial role in the construction of the flavor space
projector.  To bring them into a maximally simple form, we first note
that transformations generated by elements $h\equiv {\rm
  bdiag}(h_+,h_-)\in H$ of the sub--group $H$ of matrices
block--diagonal in advanced/retarded space leave the vacuum invariant
(up to a constant.)  Using the defining commutator relations
(\ref{cf:super_alg}) it is indeed straightforward to verify that $T_h
|\Omega\rangle = |\Omega\rangle \times {\rm sdet}(h_-)^{-1}$. Thus,
transformations $T_h$ do not change the vacuum in an `essential' way
and it is sufficient to consider the action of the coset space $G/H$
on $|\Omega\rangle$. A parameterization of individual cosets $g H \in
G/H$ optimally adjusted to our application below reads as
\begin{eqnarray*}
&&  g= \left(
    \begin{matrix}
      (1-Z\tilde Z)^{-1/2}& Z(1-\tilde Z Z)^{-1/2}\cr \tilde Z(1- Z
      \tilde Z)^{-1/2} & (1-\tilde Z Z)^{-1/2}
    \end{matrix}
\right)=\\
&&\hspace{1cm}=\left(
    \begin{matrix}
      1& Z\cr 0&1
    \end{matrix}
\right) 
\left(
\begin{matrix}
  (1-Z\tilde Z)^{+1/2}&0\cr 0& (1-\tilde Z Z)^{-1/2}
    \end{matrix}
\right)
\left(
    \begin{matrix}
      1& 0\cr \tilde Z&1
    \end{matrix}
\right).
\end{eqnarray*}
Letting the transformation $T_g$ act on the vacuum, we note that the
rightmost factor acts as the identity transformation while the
block--diagonal matrix in the middle produces a factor ${\rm
  sdet}(1-\tilde Z Z)^{1/2}$. Finally, the left factor is represented
by the linear transformation $\exp(\bar c_{+,\alpha} Z_{\alpha\alpha'}
c_{-,\alpha'})$, implying that
$$
|g\rangle = \exp(\bar c_{+,\alpha} Z_{\alpha\alpha'} c_{-,\alpha'})
|\Omega \rangle \times {\rm sdet}(1-\tilde Z Z)^{1/2},
$$
In a similar manner we obtain the conjugate action
$$
\langle g^{-1}| = {\rm sdet}(1-\tilde Z Z)^{1/2} \times \langle
\Omega |\exp(-\bar c_{-,\alpha} \tilde Z_{\alpha\alpha'}
c_{+,\alpha'}),
$$
where we noted that $g^{-1} = g\big|_{(Z,\tilde Z)\to (-Z,-\tilde
  Z)}$. We now claim that
\begin{eqnarray*}
&&{\cal P}^{II} \equiv \int_{G/H} dg\, |g\rangle \langle g |=\\  
&&\hspace{1cm}= \int d_\mu(Z,\tilde Z) 
 e^{\bar c_{+,\alpha} Z_{\alpha\alpha'}
c_{-,\alpha'}} |\Omega \rangle  \langle \Omega |e^{-\bar
c_{-,\alpha} \tilde Z_{\alpha\alpha'}
c_{+,\alpha'}},
\end{eqnarray*}
where $d_\mu (Z,\tilde Z)\equiv d(Z,\tilde Z)\, {\rm sdet}(1-Z\tilde
Z)$, and the flat measure $d(Z,\tilde Z)$is the invariant measure on
$G/K$ [] is another representation of the projector onto the flavor
space.  To prove this statement, we first note that $|g\rangle$ is an
element of the flavor space (is orthogonal to all non--flavor space
components contributing to a general Fock space state.) Second, ${\cal
  P}^{II}$ commutes with all transformations $T_g$:
\begin{eqnarray*}
  &&T_g {\cal P}^{II} = \int_{G/H} dg'\,  T_g | g'\rangle \langle
  g^{\prime -1} |=\\
&&\hspace{.2cm}=\int_{G/H} dg'\,  | gg'\rangle \langle
  g^{\prime -1} |=
\int_{G/H} dg'\,  | gg'\rangle \langle (gg')^{-1}g|
=\\
&&\hspace{1cm}=\int_{G/H} dg'\,  | g'\rangle \langle g^{\prime -1}
  |T_{g}=
{\cal P}^{II} T_g.
\end{eqnarray*}
Since the representation $g\mapsto T_g$ acts irreducibly in flavor
space, Schur's lemma implies that ${\cal P}^{II}$ must be proportional
to the unit matrix in that space. Finally, the unit normalization of
${\cal P}^{II}$ can be shown by computing the overlap $\langle \Omega
|{\cal P}^{II} |\Omega\rangle = \int d_\mu (Z,\tilde Z)\times 1 =1$,
where the last equality follows from the supersymmetry of the
integrand \cite{Zirnbauer}.

We now have everything in store to prove \eqref{eq:CF_trafo}.  Noting
that $|\nu_+,\nu_-\rangle \equiv \exp(\bar c_{+\alpha} \nu_{+\alpha} +
\bar \nu_{-\alpha} c_{-\alpha} )|\Omega\rangle$ and $\langle
\eta_+,\eta_-|\equiv \langle \Omega |\exp(\eta_{+\alpha} c_{+\alpha} +
\bar c_{-\alpha} \eta_{-\alpha})$ are coherent states, i.e.
\begin{eqnarray*}
&&  c_{+\alpha} |\nu_+,\nu_-\rangle = +\nu_{+\alpha}
|\nu_+,\nu_-\rangle,\\
&&\bar c_{-\alpha} |\nu_+,\nu_-\rangle = 
-\nu_{-\alpha} |\nu_+,\nu_-\rangle,\\
&&\\
&&\langle \eta_+,\eta_-| \bar c_{+\alpha} =
\langle \eta_+,\eta_- |\, (+\eta_{+\alpha}),\\
&&\langle \eta_+,\eta_-|  c_{-\alpha} =
\langle \eta_+,\eta_- |\, (-\eta_{-\alpha})
\end{eqnarray*}
the color flavor--transformation may be proven by mapping the l.h.s.
of Eq. \eqref{eq:CF_trafo} to a Fock--space matrix element and using
the equality of the two projector representations derived above:
\begin{widetext}
  \begin{eqnarray*}
 &&  \int \frac{d\phi}{2\pi}\,
  e^{
    \eta_{+}^Te^{i\phi}\nu_{+} + 
    \nu_{-}^Te^{-i\phi}\eta_{-}
  }
  =    \int \frac{d\phi}{2\pi}\,\langle \Omega |e^{\eta_{+\alpha}
  c_{+\alpha} - \bar c_{-\alpha} \eta_{-\alpha} }\,
e^{\bar c_{+\alpha} e^{i\phi} \nu_{+\alpha} + \nu_{-\alpha} e^{-i\phi}
  c_{-\alpha}}|\Omega\rangle=\\
&&\hspace{1cm}=\langle \eta+,-\eta_- |\,
{\cal P}^I |\nu_+,\nu_-\rangle =\langle \eta_+,-\eta_-|\,
{\cal P}^{II} |\nu_+,\nu_-\rangle=\\
&&\hspace{1cm}=
\int d_\mu( Z,\tilde Z)\langle \eta_+,-\eta_-| e^{\bar c_{+,\alpha} Z_{\alpha\alpha'}
c_{-,\alpha'}} |\Omega \rangle  \langle \Omega |e^{-\bar
c_{-,\alpha} \tilde Z_{\alpha\alpha'}
c_{+,\alpha'}} |\nu_+,\nu_-\rangle=\\
&&\hspace{1cm}=\int d_\mu( Z,\tilde Z)\, e^{ \eta_{+,\alpha} Z_{\alpha\alpha'}
\eta_{-,\alpha'}+
\nu_{-,\alpha} \tilde Z_{\alpha\alpha'}
\nu_{+,\alpha'}}\, \langle \eta_+,-\eta_-  |\Omega \rangle  \langle
  \Omega  |\nu_+,\nu_-\rangle=\\
&&=\int d_\mu( Z,\tilde Z)\, e^{ \eta^T_{+} Z
\eta_{-}+
\nu_{-}^T \tilde Z
\nu_{+}}.
  \end{eqnarray*}
\end{widetext}

\begin{acknowledgments}
  We have enjoyed fruitful discussions with Fritz Haake, Sebastian
  M\"uller, Stefan Heusler, and Peter Braun. This work has been
  supported by SFB/TR12 of the Deutsche Forschungsgemeinschaft.
  SG thanks for support by the Minerva Foundation.
\end{acknowledgments}


\begin{thebibliography}{}
\bibitem{bondscattering} The quantum map is also
  known as the bond scattering matrix.
\bibitem{double_dim} The doubling of dimensions at this point is 
  convenient for the later introduction of a time-reversal space
  for the fields. It can only be justified a posteriori.
\bibitem{Gaussian}In the Gaussian approximation the integration over
  $Z_\mathbf{bb}$ is not restricted.
\bibitem{stars} In a star graph with $B$ bonds $B$ peripheral vertices
  of unit valency are connected to one central vertex of valency $B$.
\end{thebibliography}
\end{document}